# Neutron spectroscopic factors of $^{55}$Ni hole-states from (p,d) transfer reactions

A. Sanetullaev[1+], M.B. Tsang(曾敏兒)[1*], W.G. Lynch(連致標)[1], Jenny Lee(李曉菁)[1], D. Bazin[1], K.P. Chan[1,2] (陳家鵬), D. Coupland[1], V. Henzl[1], D. Henzlova[1], M. Kilburn[1], A.M. Rogers[1], Z.Y. Sun(孙志宇)[1,3], M. Youngs[1], R.J. Charity[4], L.G. Sobotka[4], M. Famiano[5], S. Hudan[6], D. Shapira[7], W.A. Peters[7], C. Barbieri[8], M. Hjorth-Jensen[1,9], M. Horoi[10], T. Otsuka[11], T. Suzuki[12], Y. Utsuno[13]

[1]National Superconducting Cyclotron Laboratory & Department of Physics and Astronomy, Michigan State University, East Lansing, Michigan 48864, USA
[2]Physics Department, Hong Kong Chinese University, Shatin, Kong Kong, China
[3]Institute of Modern Physics, CAS, Lanzhou 730000, China
[4]Department of Chemistry, Washington University, St. Louis, MO 63130, USA
[5]Department of Physics, Western Michigan University, Kalamazoo, MI 49008, USA
[6]Department of Chemistry, Indiana University, Bloomington, IN 47405, USA
[7]Oak Ridge National Laboratory, Oak Ridge, TN 37831, USA
[8]Department of Physics, University of Surrey, Guildford GU2 7XH, United Kingdom
[9]Department of Physics and Center of Mathematics for Applications, University of Oslo, N-0316 Oslo, Norway
[10]Department of Physics, Central Michigan University, Mount Pleasant, MI 48859, USA
[11]Department of Physics, University of Tokyo, Tokyo, Japan
[12]Department of Physics, Nihon University, Tokyo, Japan
[13]Advanced Science Research Center, Japan Atomic Energy Agency, Tokai, Ibaraki 319-1195, Japan

**Abstract**

Spectroscopic information has been extracted on the hole-states of $^{55}$Ni, the least known of the quartet of nuclei ($^{55}$Ni, $^{57}$Ni, $^{55}$Co and $^{57}$Co), one neutron away from $^{56}$Ni, the N=Z=28 double magic nucleus. Using the $^{1}$H($^{56}$Ni,d)$^{55}$Ni transfer reaction in inverse kinematics, neutron spectroscopic factors, spins and parities have been extracted for the $f_{7/2}$, $p_{3/2}$ and the $s_{1/2}$ hole-states of $^{55}$Ni. This new data provides a benchmark for large basis calculations that include nucleonic orbits in both the sd and pf shells. State of the art calculations have been performed to describe the excitation energies and spectroscopic factors of the $s_{1/2}$ hole-state below Fermi energy.

+Present address: TRIUMF, Vancouver, BC Canada
* Corresponding author: tsang@nscl.msu.edu

Doubly-magic nuclei, with both the proton number (Z) and the neutron number (N) corresponding to closed shells, have played a simplifying role in nuclear structure. Low lying states in somewhat heavier nuclei are often approximated by ignoring the closed shell single particle orbits and considering only the occupied nucleon orbits in higher lying "particle" states in the next major shell. Similarly, low lying states in lighter nuclei are often approximated by considering only unoccupied nucleon orbits ("holes") within the doubly closed shell. In this approximation, effective operators on the particle- or hole- states provide a means to model the nucleonic wavefunctions in the closed shell "core". This simplification works extremely well in the vicinity of the doubly-magic closed sd shell nucleus $^{40}$Ca.

$^{56}$Ni is the first doubly-magic N=Z nucleus, beyond $^{40}$Ca. Produced abundantly in stellar reactions, it plays important roles in many astrophysical processes [1,2]. Despite its importance, its six-day half-life hindered key investigations of $^{56}$Ni and its neighbors until radioactive beams became available. Consequently, there are very little data for $^{55}$Ni with one neutron less than $^{56}$Ni. Only six out of 22 known states below 7 MeV in this nucleus have tentative spins and parities [3]．The only firmly established spin and parity in $^{55}$Ni is the ground state with spin and parity of $7/2^-$ [3,4]. In addition to clarify the extent to which the $^{56}$Ni calculations can be regarded as a closed $1f_{7/2}$ core, measurements of hole-states in $^{55}$Ni provide the excitation energies of filled $1d_{3/2}$ and $2s_{1/2}$ levels below the Fermi energy. The properties and locations of these positive parity excited "hole-" states in $^{55}$Ni would give the energy separations between major shells, which may be linked to saturation properties and to the possible role of three-body forces [5-8].

In this paper, we report the first spectroscopic factor measurements of neutron hole states in $^{55}$Ni. We find that their successful interpretation requires very large basis state-of-the-art calculations involving orbits in both the fp and sd shells. The result may bridge our knowledge gap between the successful interactions for the sd shell and the less successful cousins interactions that describe the pf orbits. Comparing to ab-initio calculations we further find that the current two-nucleon only chiral interactions overestimate the inter-shell gaps.

The experiment was performed at the Coupled Cyclotron Facility of Michigan State University. There, a 140 MeV/nucleon primary beam of $^{58}$Ni bombarded a 1269 mg/cm$^2$ beryllium production target. The A1900 fragment separator filtered $^{56}$Ni nuclei from the resulting fragmentation products to a purity of 72% and degraded their energies to 37 MeV/nucleon. Details on the identification of the resulting $^{56}$Ni beam can be found in Ref [9, 10].

This secondary $^{56}$Ni beam impinged upon a 9.6 mg/cm$^2$ polyethylene (CH$_2$)$_n$ target. To correct for large size (11mm x 17 mm FWHM) and angular divergence (~1.4°) of this secondary beam, two microchannel plate (MCP) detection systems, 10 cm and 60 cm upstream of target, determined the positions and angles of the incoming projectile on the target [9,10]. This tracking system can achieve a position resolution of ~1 mm for beam intensities up to 5×10⁵ pps [9]. The MCP closest to the target monitored the absolute beam intensity throughout the experiment.

Sixteen telescopes of the High Resolution Array, (HiRA) [11], measured the energies and angles of the deuterons and thereby identified the $^{55}$Ni states populated by the $^{1}$H($^{56}$Ni, d)$^{55}$Ni reaction. These telescopes were located 35 cm from the target where they covered polar angles of 4°- 45° in the laboratory frame. Each telescope contained one 65 μm thick ΔE and a 1500 μm thick E silicon strip detector that were backed by four 3.9 cm thick CsI(Tl) crystals. The 32 horizontal and 32 vertical strips in the E detector effectively subdivided each telescope into 1024 2mm x 2mm pixels, each with an angular resolution of about ±0.16°. Each ΔE detector has 32 strips in the front which are aligned with the vertical strips of the E detector. The Laser Based Alignment System (LBAS) [12] provided the locations of each pixel relative to the target to an accuracy of 0.3 mm.

The HiRA, MCP and reaction targets were placed in the S800 scattering chamber in front of the S800 spectrometer [13,14], which detected the $^{55}$Ni residues. This kinematically complete measurement of deuterons and $^{55}$Ni ensures that both particles originate from the $^{1}$H($^{56}$Ni, d)$^{55}$Ni reaction. Measurements with a $^{12}$C target verified that the $^{12}$C nuclei in the polyethene target contributed negligibly to the deuteron cross-sections investigated in this paper.

The left panel of Fig. 1 shows the 2D spectrum of the laboratory angle vs. laboratory energy of the emitted deuteron in coincidence with $^{55}$Ni detected in the S800 focal plane, obtained without benefit of the MCP beam tracking detectors. Two bands corresponding to the ground state of $^{55}$Ni and an excited state around 3.18 MeV can be identified. Deuterons with $E_{Lab}$ < 22 MeV stop in the silicon E detector and deuterons with 22 MeV < $E_{Lab}$ < 155 MeV stop in the CsI(Tl) crystals. The maximum laboratory angle of the deuterons in this reaction is approximately 32°; there the energy for ground state band increases rapidly with center of mass scattering angle.

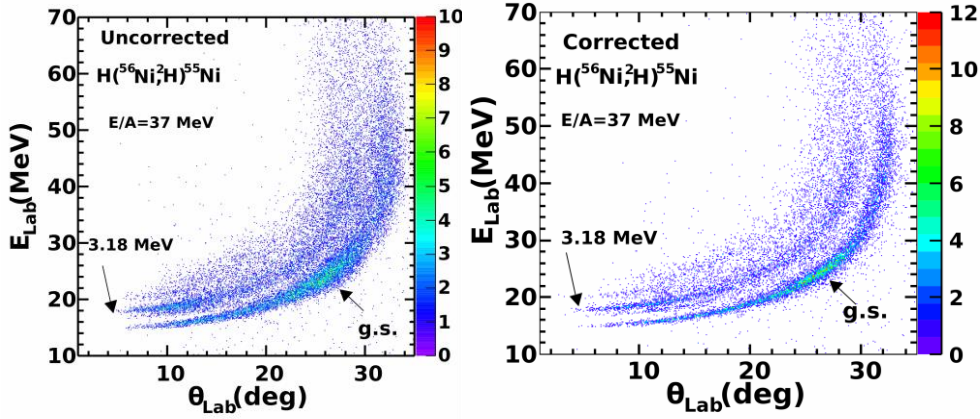

Figure 1: (Color online) Left panel: 2D plot of angle vs. energy for the deuterons detected in the HiRA. Right panel: same as left panel after correcting the beam position and angle using micro-channel plate tracking detectors.

The right panel of Fig. 1, shows the improved resolution obtained by correcting the deuteron scattering angle for the position and angle of each beam particle at the target using the MCP beam trackers [9]. At forward angles around 15°, the first excited state of $^{55}$Ni at 2.09 MeV can be clearly seen between the stronger bands corresponding to the ground state and the 3.18 MeV excited state. Near the maximum deuteron emission angle of ~30° in the laboratory, (which corresponds to roughly 40° in the center of mass,) the excitation energy resolution worsens due to a number of factors, including target thickness. As cross sections at these larger angles are more sensitive to ambiguities in optical model potentials, we focus on data taken at angles up to 30 deg in the laboratory frame.

Fig. 2 shows the reconstructed excitation energy of $^{55}$Ni at six angles in the center of mass frame. For best resolution, we choose forward angles where the deuteron stops in the Silicon E detectors. Three peaks, corresponding to the ground state, first-excited state at 2.09 MeV, and 3.18 MeV state, can be clearly identified in each angular range in the figure. The energy resolutions of the ground and 3.18 MeV peaks are both about 550 keV FWHM. This energy resolution agrees with GEANT4 simulations taking into account the energy and angular straggling in the target and detector resolutions [15]. There is additional yield at excitation energies of approximately 3.75 MeV, but the background at this excitation energy is uncertain. The worsening energy resolution at large angles, $\theta_{LAB} > 17°$ does not allow the extraction of precise differential cross sections for the 3.75 MeV state.

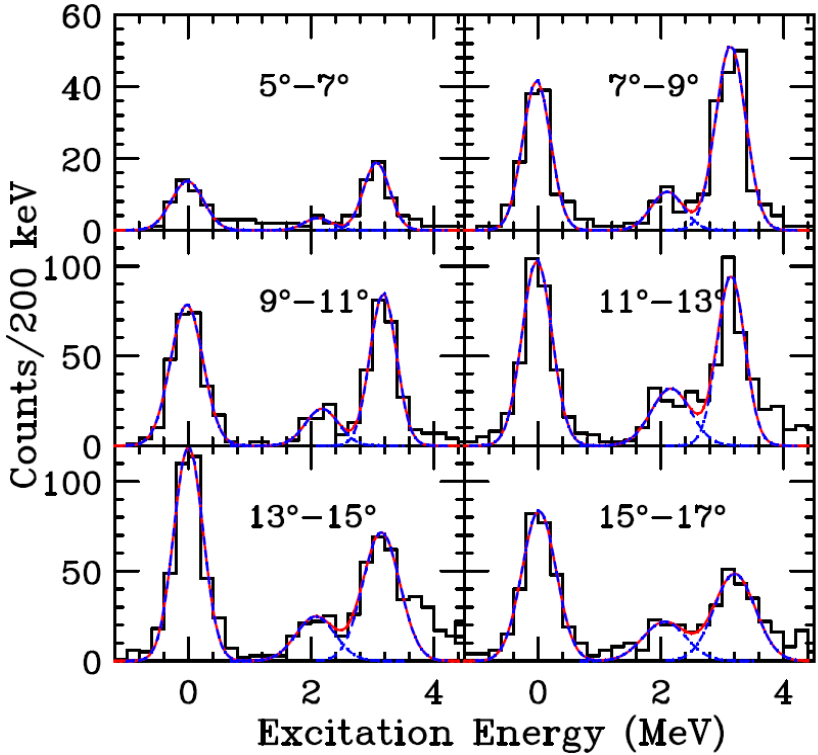

Figure 2: (Color online) Excitation energy of $^{55}$Ni constructed from the emitted deuterons in the $^{1}$H($^{56}$Ni, d)$^{55}$Ni reaction. Dashed curves are Gaussian fits for individual states and the solid curve is the sum of all the fits.

At the most forward angles, (top left panel), the background from random is higher due to forward peaking of most emitted particles. The total coincident deuteron counts detected are lower because of the low geometrical coverage of the HiRA array at these angles. By using Gaussian fits, we are able to extract the counts for all three states identified here in these six angular bins. At larger angles, contributions from the 2.09 MeV state become insignificant.

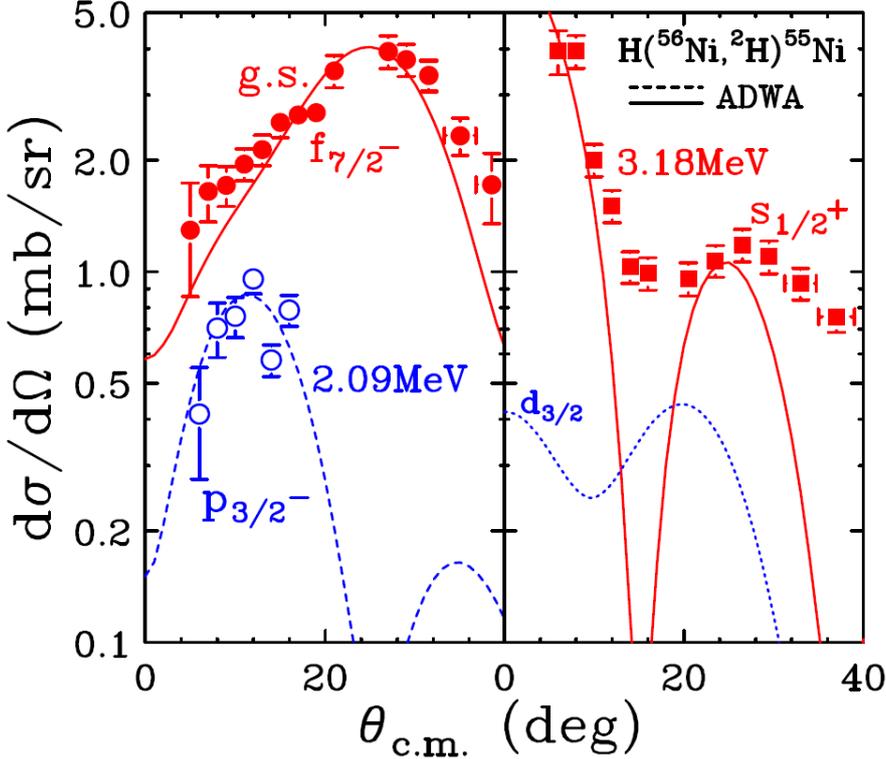

Figure 3: (color online) Deuterons angular distributions for different states of $^{55}$Ni. Curves are calculations from ADWA reaction model for individual states as indicated on the figure. Except for the $d_{3/2}$ state which assumes a normalization value of 1, the normalization constants of the other curves yield the spectroscopic values as described in the text.

The angular momentum of the states is extracted from the measured angular distributions shown in Fig. 3. Definitive spin and parities of states are usually obtained using polarized beam which is difficult to do in inverse kinematics. However, the help of the systematics of the N=27 isotones, we can confirm the tentative assignment of the spins and parities for the $f_{7/2}$, $p_{3/2}$ and the $s_{1/2}$ hole-states of $^{55}$Ni. The left panel shows the differential cross-sections for the ground state (solid points) and the first-excited state (open circles). The large separation of the ground-state from the excited-states allows unambiguous extraction of the ground-state peak at all angles. The ground-state angular distribution peaks around 22° with a characteristic shape of $l=3$ transfer. Using the algorithm developed in ref. [16], we extract the neutron ground state spectroscopic factor of $^{56}$Ni(g.s) to be 6.7±0.7, close to the independent particle model value of 8 for a closed $f_{7/2}$ shell. The solid curve is a calculation with the Adiabatic Distorted Wave Approximation (ADWA) [17] using TWOFNR code [18] with the calculated cross section multiplied by 6.7. The calculation describes the shape of the angular distribution reasonably well.

The angular distribution of the weak peak at 2.09 MeV is shown as the open circles in the left panel of Fig. 3. Unlike the ground state, the angular distribution of the 2.09 MeV state peaks around 10° and can be described very well by the angular distributions of a $p_{3/2}$ state. Taking the one-step direct reaction mechanism of the ADWA formalism to be correct implies that the ground state of $^{56}$Ni contains a small admixture of a valence neutron in the $p_{3/2}$ orbit, which lies above the N=28 shell gap. The small spectroscopic factor of 0.14±0.03, extracted for this state, may indicate a dominant 2-hole-1-particle characteristic and it is confirmed by the theoretical calculations below.

Finally, the angular distribution of the third state at 3.18 MeV is plotted in the right panel of Fig. 3. The differential cross-sections of the first six points are extracted from the Gaussian fits shown as dashed curves in Fig. 2. Beyond 19°, the cross-sections of the 2.09 MeV state drop to negligible levels and single Gaussian fits are sufficient to extract the differential cross-sections. The forward peaking of the angular distribution is consistent with $l=0$ orbit suggesting that this is the $s_{1/2}$ hole-state below the Fermi-energy in the 1d2s major shell which lies below the $1f_{7/2}$ major shell. The dominance of the $s_{1/2}$ state is evident in Fig. 2 and the existence of such a state is consistent with the systematic of the N=27 isotones [19]. The solid curve is the ADWA calculation assuming $^1$H($^{56}$Ni(g.s), d)$^{55}$Ni(3.18, $s_{1/2}$) transition consistent with removing a valence neutron from the $s_{1/2}$ orbit in $^{56}$Ni. Due to the finite angular resolution of the detector, the predicted deep valley around 18° cannot be measured resulting in a shallow valley in the angular distribution. Around this angle, deuteron with energy > 22 MeV penetrate the E silicon detectors and enter the CsI(Tl) crystals which has worse resolution. The extracted spectroscopic factor is 1.0±0.2, exhausting roughly 50% of the total strength expected in the independent particle model.

Beyond 9°, we observe some excess strength at E*>3.5 MeV in Fig. 2. However, the background does not allow an independent confirmation of a peak at 3.75 MeV. The compilation of levels for the $^{55}$Ni nucleus by the National Nuclear Data Center (NNDC) [20, 3, 4] assigns a tentative spin of (3/2$^+$) for the 3.75 MeV excited state. The dotted curve inside the right panel of Fig. 3 represents the calculations of the $d_{3/2}$ state using ADWA model assuming a normalization factor of 1. From Fig. 3, we estimate that the $d_{3/2}$ hole-state cross section would be a factor of 5-10 times smaller than that of the $s_{1/2}$ state at $\theta_{LAB}$<15°, roughly consistent with the observed strength. Unfortunately, at larger angles where the $d_{3/2}$ state should have a larger cross section, the deuterons punch into the CsI(Tl) detectors and both the energy and kinematic angular resolution is not good enough to make definitive conclusions.

The shell-model excitation energies of low-lying $f_{7/2}$ and $p_{3/2}$ states can be calculated relatively accurately in the pf valence space (PF) using the G-matrix (GX) based effective interaction GXPF1A of Refs. [21,22]. To describe the $s_{1/2}$ states, however, requires a valence space that mixes relevant sd (SD) with pf orbitals. Results in Table I are obtained from the SDPFM interaction of Refs. [23,24], which has been modified here by freezing the $d_{5/2}$ orbital in the core and adjusting the single particles energies and the sd-pf monopoles to describe the low-lying states in $^{55}$Ni. The results are shown by the columns labels "SM1" in Table I and predict the properties of deep-hole states in the sd orbitals below the Fermi surface reasonably well. Note that absolute spectroscopic factors are in general overestimated in small valence spaces [6,25]. However it is still possible to provide sensible predictions about their relative values from shell model calculations [6,26]. This is important to identify dominant single particle states. To better test our conclusions, we also activate all the sd and pf orbits as valence shells but truncate many-body states appropriately. Specifically, we assume $1/2^+_1$ and $3/2^+_1$ to be dominantly one-hole states, limit the number of nucleons excited from the sd shell to the pf shell to one, and restrict the number of nucleons in the upper pf orbits ($p_{3/2}$, $f_{5/2}$ and $p_{1/2}$) to be equal to or less than six. We use this truncation with the SDPF-MU interaction of Ref. [26], which was obtained similarly to SDPFM but adding universal monopole-based corrections for the interaction between the sd and pf orbits [27]. This approach provided reasonable agreement with experimental energy levels and the spectroscopic factors, as shown by columns labeled "SM2" in Table I.

| State | E*_expt (MeV) | E*_SM1 (MeV) | E*_SM2 (MeV) | E*_SCGF (MeV) | SF_expt | SF_SM1 | SF_SM2 | SF_SCGF |
|---|---|---|---|---|---|---|---|---|
| $f_{7/2}$ | 0 | 0 | 0 | 0 | 6.7±0.7 | 6.75 | 6.94 | 5.78 |
| $p_{3/2}$ | 2.09 | 1.895 | 2.281 | 4.34 | 0.19±0.03 | 0.189 | 0.13 | 0.06 |
| $s_{1/2}$ | 3.18 | 3.039 | 3.755 | 11.48 | 1.0±0.2 | 1.57 | 1.21 | 1.10 |
| $d_{3/2}$ | (3.752) | 3.309 | 4.453 | 12.47 | NA | 2.88 | 1.68 | 2.16 |

Table I: Experimental and calculated information on the $f_{7/2}$, $p_{3/2}$ and $s_{1/2}$ states. The notation of "SM1", "SM2", and "SCGF" refers to the shell-model calculations using a modified SDPFM interaction, those with the SDPF-MU interaction, and those with self-consistent Green's functions theory respectively.

Calculations were also performed with ab-initio self-consistent Green's functions (SCGF) theory [28], which can estimate separation energies and spectroscopic factors of quasi-hole peaks away from the major valence shell. The results, based on the two-body chiral next-to-next-to-next-to-leading order (N3LO) interaction [29], puts the $p_{3/2}$ state at 4.3 MeV with a weak spectroscopic factor [28]. However, the dominant $s_{1/2}$ and $d_{3/2}$ states are predicted being bound too deeply, with excitation energies around 12 MeV rather than the observed values. Together with the dominant peaks shown in Table I, other satellite states are predicted to spread between 11 and 18 MeV.

Nonetheless, the spectroscopic factors predicted by SCGF appear in rather good agreement with the observation, though.

Overall, the results of Table I confirm the interpretation of the 3.18 MeV excited state as the dominant $s_{1/2}$ quasi-hole peak, and support the theoretical predictions that the $1d_{3/2}$ orbit is about 1 MeV higher in excitation energy and inverted with respect to the $1s_{1/2}$. The too high excitation energies of all the sd states obtained by SCGF theory indicates that the current two-body N3LO interaction tends to overestimate gaps between major shells. It would be interesting to perform these calculations that include three-body forces.

**Summary and Conclusion**

In the present work, the deuteron cross-sections are extracted from a kinematically-complete experiment $^1$H($^{56}$Ni, d)$^{55}$Ni using a high resolution array, micro channel plate tracking detectors and the S800 spectrometer. The good angular and energy resolution of the data allows us to extract the spectroscopic information for three low-lying states of $^{55}$Ni. In addition, we also extract the neutron spectroscopic factors of these states. In general, spectroscopic factors of dominant peaks of 50-69% suggest that $^{56}$Ni can be seen as a closed shell. This view is supported by the present calculations and data. The data also indicates that the separation between sd and pf shells is of 3-4 MeV. Unlike the case for hole states in $^{39}$Ca made by removing one neutron from doubly closed shell $^{40}$Ca, the description of the low lying structure of $^{55}$Ni requires a model space that spans both the sd and fp major shells. To reproduce the excitation energies and spectroscopic factors, the state-of-the-art shell model calculations and improved effective interactions are needed. Self-consistent Green's functions theory cannot explain the correct gaps between sd and pf orbits with chiral Hamiltonians. Since studies of three-nucleon forces in finite nuclei have so far focused mainly on the evolutions within single shells [7] or within isotope chains [30,31], it might be important for future studies to further compare theory and data for isotone chains and other Ni isotopes, moving from stability to instability.

**Acknowledgements:**

The authors would like to thank Professor B.A. Brown for stimulating discussions and Professor J. Tostevin for the use of the programs TWOFNR. This work is supported by the National Science Foundation under Grant No. PHY- 1102511. M.H. acknowledges U.S. NSF Grant No. PHY-1068217. C.B. acknowledges the UK STFC Grant No. ST/J000051/1.

**References:**

[1] M.B. Aufderheide, G.E. Brown, T.T.S. Kuo, D.B. Stout, P. Vogel, Astrophys. J. **362** (1990) 241.

[2] R. K. Wallace and S. E. Woosley, Astrophys. J., Suppl. Ser. **45** (1981) 389.

[3] Huo Junde, Nuclear Data Sheets, **109** (2008) 787.

[4] D. Mueller, E. Kashy, and W. Benenson, Phys. Rev. C **15** (1977) 1282.

[5] K. Hebeler et al., Phys. Rev. C **83** (2011) 031301(R).

[6] V. Somà and P. Bożek., Phys. Rev. C **78** (2008) 054003.

[7] A. P. Zuker, Phys. Rev. Lett. **90** (2003) 042502.

[8] T. Otsuka, T. Suzuki, J. D. Holt, A. Schwenk, and Y. Akaishi, Phys. Rev. Lett. **105** (2010) 032501

[9] A. M. Rogers et al., arXiv:1309.2745 (2013).

[10] A. Sanetullaev, PhD thesis, Michigan State University (2011).

[11] M.S. Wallace et al., Nucl. Instr. and Meth. A**583** (2007) 302.

[12] A.M. Rogers et al., Nucl. Instr. and Meth. A **707** (2013) 64.

[13] J. Yurkon et al., Nucl. Instr. and Meth. A **422** (1999) 291.

[14] D. Bazin et al., Nucl. Instr. and Meth. B **204** (2003) 629.

[15] Jenny Lee, et al., Phys. Rev. C 83 (2011) 014606.

[16] Jenny Lee, M.B. Tsang, W. G. Lynch, Phys. Rev. C **75** (2007) 064320.

[17] R.C. Johnson and P.J.R. Soper, Phys. Rev. C**1** (1970) 976.

[18] M. Igarashi et al., Computer Program TWOFNR (Surrey University version).

[19] F. Lu et al., Phys. Rev. C **88** (2013) 017604.

[20] National Nuclear Data Center, http://www.nndc.bnl.gov/

[21] M. Honma, T. Otsuka, B.A. Brown, and T. Mizusaki, Phys. Rev. C **69**, (2004) 034335.

[22] M. Honma, T. Otsuka, B.A. Brown, and T. Mizusaki, Eur, Phys, J. A **25**, Suppl.1, 499 (2005).

[23] Y. Utsuno, T. Otsuka, T. Mizusaki, and M. Honma, Phys. Rev. C **60** (1999) 054315.

[24] E.S. Diffenderfer et al, Phys. Rev. C **85** (2012) 034311.

[25] C. Barbieri, Phys. Rev. Lett. **103** (2009) 202502.

[26] Y. Utsuno, T. Otsuka, B. A. Brown, M. Honma, T. Mizusaki, and N. Shimizu, Phys. Rev. C **86** (2012) 051301(R).

[27] T. Otsuka, T. Suzuki, M. Honma, Y. Utsuno, N. Tsunoda, K. Tsukiyama, and M. Hjorth-Jensen, Phys. Rev. Lett. **104** (2010) 012501.

[28] C. Barbieri, M. Hjorth-Jensen, Phys. Rev. C **79** (2009) 064313.

[29] R. Machleidt and D. R. Entem, Phys. Rep. **503** (2011) 1.

[30] J. Lee, M.B. Tsang, W. G. Lynch, M. Horoi, S. C. Su, Phys. Rev. C **79** (2009) 054611.

[31] A. Cipollone, C. Barbieri, and P. Navrátil, Phys. Rev. Lett. **111** (2013) 062501.